\documentclass[numberedappendix,apj,twocolumn]{emulateapj}

\newcommand{\zy}{$z_{850} - Y_{105}$}
\newcommand{\yj}{$Y_{105} - J_{125}$}
\newcommand{\jh}{$J_{125} - H_{160}$}
\newcommand{\lya}{Ly$\alpha$}

\newcommand{\zdrop}{$z_{850}$-dropout}

\newcommand{\bFilter}{$B_{435}$}
\newcommand{\vFilter}{$V_{606}$}
\newcommand{\iFilter}{$i_{775}$}
\newcommand{\zFilter}{$z_{850}$}
\newcommand{\yFilter}{$Y_{105}$}
\newcommand{\jFilter}{$J_{125}$}

\newcommand{\hFilter}{$H_{160}$}

\shorttitle{First Epoch HUDF09: $z_{850}$-dropouts}
\shortauthors{Oesch et al.}

\begin{document}

\title{$z\sim7$ Galaxies in the HUDF: First Epoch WFC3/IR Results
\altaffilmark{1}}

\altaffiltext{1}{Based on data obtained with the \textit{Hubble Space Telescope} operated by AURA, Inc. for NASA under contract NAS5-26555. }

\author{P. A. Oesch\altaffilmark{2},
R. J. Bouwens\altaffilmark{3,4}, 
G. D. Illingworth\altaffilmark{3}, 
C. M. Carollo\altaffilmark{2}, 
M. Franx\altaffilmark{4}, 
I. Labb\'{e}\altaffilmark{5}, 
D. Magee\altaffilmark{3}, 
M. Stiavelli\altaffilmark{6},
M. Trenti\altaffilmark{7}, 
P. G. van Dokkum\altaffilmark{8}
}

\altaffiltext{2}{Institute for Astronomy, ETH Zurich, 8092 Zurich, Switzerland; poesch@phys.ethz.ch}
\altaffiltext{3}{UCO/Lick Observatory, University of California, Santa Cruz, CA 95064}
\altaffiltext{4}{Leiden Observatory, Leiden University, NL-2300 RA Leiden, Netherlands}
\altaffiltext{5}{Carnegie Observatories, Pasadena, CA 91101, Hubble Fellow}
\altaffiltext{6}{Space Telescope Science Institute, Baltimore, MD 21218, United States}
\altaffiltext{7}{University of Colorado, Center for Astrophysics and Space Astronomy,
389-UCB, Boulder, CO 80309, USA}
\altaffiltext{8}{Department of Astronomy, Yale University, New Haven, CT 06520}

\begin{abstract}
We present a sample of 16 robust $z\sim7$ $z_{850}$-drop galaxies detected by the newly installed WFC3/IR camera on the Hubble Space Telescope. Our analysis is based on the first epoch data of the HUDF09 program covering the Hubble Ultra Deep Field with 60 orbits of \yFilter, \jFilter, and \hFilter\ observations. These remarkable data cover 4.7 arcmin$^2$ and are the deepest NIR images ever taken, reaching to $\sim 29$ mag AB ($5\sigma$). The 16 $z\sim6.5-7.5$ galaxies have been identified based on the Lyman Break technique utilizing (\zy) vs. (\yj) colors. They have magnitudes \jFilter$ = 26.0-29.0$ (AB), an average apparent half-light radius of $\sim0.16$ arcsec ($\lesssim1$ kpc), and show very blue colors (some even $\beta\lesssim-2.5$), in particular at low luminosities. The WFC3/IR data confirms previous NICMOS detections indicating that the dropout selection at $z\sim7$ is very reliable.
Our data allow a first determination of the faint end slope of the $z\sim7$ luminosity function, reaching down to $M_{UV}\sim-18$, a full magnitude fainter than previous measurements. 
When fixing $\phi_*=1.4\times10^{-3}$ Mpc$^{-3}$mag$^{-1}$ to the value previously measured at $z\sim6$, we find a best-fit value of $\alpha=-1.77\pm0.20$, with a characteristic luminosity of $M_*=-19.91\pm0.09$.
This steep slope is similar to what is seen at $z\sim2-6$ and indicates that low luminosity galaxies could potentially provide adequate flux to reionize the universe. The remarkable depth and resolution of these new images provide insights into the coming power of JWST.

\end{abstract}


\section{Introduction}
While great progress has been made in the study of galaxies up to $z\sim6$, even after more than a decade of Lyman break galaxy surveys, the samples of UV selected candidate $z\gtrsim7$ galaxies are still very modest \citep[e.g.][]{bouw04b,yan04,bouw08,brad08,rich08,oesch09,zheng09,ouchi09,castellano09}. 
Due to their small number, the $z\gtrsim7$ luminosity function (LF), and in particular its faint end slope, is still very uncertain. An accurate measurement of this is of fundamental importance, however, not only for the early stages of galaxy formation, but also for the understanding of the epoch of reionization and the role of galaxies therein.

For sources of such early cosmic times the inter-galactic medium absorbs most of the light blueward of \lya, which is redshifted into the near infrared (NIR), where the sensitivities and fields of view of the detectors have lagged behind the optical CCD cameras. 
Extremely deep NIR imaging is needed to detect these galaxies. Moreover, due to limited spatial resolution of earlier IR cameras, it has not been possible so far to measure other basic quantities of $z\sim7$ galaxies such as sizes.

With the installation of the Wide Field Camera 3 (WFC3) in Service Mission 4 of the Hubble Space Telescope (HST) a new era of high redshift galaxy surveys has begun. With its large area, high sensitivity and small pixel size it allows us to probe faint objects much more efficiently than HST's previous NIR camera NICMOS (Near Infrared Camera and Multi-Object Spectrometer). Additionally, WFC3's improved filter set over NICMOS helps to exclude contaminating interloper objects and to estimate more precise photometric redshifts for these sources. Furthermore, it allows for the first time the detection of galaxies at $z\sim8$, which we report on in an accompanying paper \citep[][submitted]{bouw09b}.

In this paper we present early results on our search for $z\sim7$ galaxies from the first epoch data of the HUDF09 survey. In \S \ref{sec:data} we describe the data, the $z \sim 7$ candidate selection and its efficiency. In \S \ref{sec:LF} we present new constraints on the LBG LF at $z\sim7$.
We adopt $\Omega_M=0.3, \Omega_\Lambda=0.7, H_0=70$ kms$^{-1}$Mpc$^{-1}$, i.e. $h=0.7$. Magnitudes are given in the AB system \citep{okeg83}.

\begin{deluxetable*}{lcccccccc}
\tablecaption{Photometry (AB mag) of the $z\sim7$ $z_{850}$-dropout sources\tablenotemark{*}\label{tab:phot}}
\tablewidth{0pt}
\tablecolumns{9}
%\rotate
\tablehead{\colhead{ID} & $\alpha$ & $\delta$ &\colhead{\jFilter}  &\colhead{\zy} & \colhead{\yj}& \colhead{\jh} & 
   \colhead{S/N$_{J125}$}  &\colhead{REF.}  }

\startdata
UDFz-42566566  &  03:32:42.56  &  -27:46:56.6  &  $25.95\pm0.04$ & $1.48\pm0.14$  &  $0.16\pm0.04$ &$-0.08\pm0.04$ & 25.8  & $1,2,3,4,5,6$    \\

UDFz-44716442  &  03:32:44.71  &  -27:46:44.2  &  $26.90\pm0.10$ & $>2.43$  &  $0.66\pm0.12$ &$-0.25\pm0.09$ & 17.2  & $\times$    \\
UDFz-38807073  &  03:32:38.80  &  -27:47:07.3  &  $26.90\pm0.06$ & $2.34\pm0.61$  &  $0.41\pm0.08$ &$0.14\pm0.06$ & 26.2  & $1,2,4,5,6,7$    \\
UDFz-39557176  &  03:32:39.55  &  -27:47:17.6  &  $27.19\pm0.08$ & $1.95\pm0.70$  &  $0.51\pm0.12$ &$0.04\pm0.09$ & 14.8  & $2,5$    \\
UDFz-42577314  &  03:32:42.57  &  -27:47:31.4  &  $27.22\pm0.20$ & $1.61\pm0.33$  &  $0.30\pm0.14$ &$-0.16\pm0.12$ & 11.1  & $2,5,6$    \\
UDFz-39586565  &  03:32:39.58  &  -27:46:56.5  &  $27.74\pm0.18$ & $0.99\pm0.27$  &  $-0.11\pm0.15$ &$-0.32\pm0.18$ & 9.3  & $-$    \\
UDFz-37228061  &  03:32:37.22  &  -27:48:06.1  &  $27.76\pm0.14$ & $>2.18$  &  $0.46\pm0.16$ &$-0.08\pm0.13$ & 11.3  & $-$    \\
UDFz-43146285  &  03:32:43.14  &  -27:46:28.5  &  $27.81\pm0.20$ & $>2.13$  &  $0.27\pm0.16$ &$0.04\pm0.14$ & 10.3  & $\times$    \\
UDFz-36777536  &  03:32:36.77  &  -27:47:53.6  &  $27.83\pm0.16$ & $1.06\pm0.23$  &  $-0.23\pm0.11$ &$-0.13\pm0.13$ & 13.5  & $-$    \\
UDFz-37446513  &  03:32:37.44  &  -27:46:51.3  &  $27.86\pm0.16$ & $1.05\pm0.27$  &  $-0.11\pm0.13$ &$-0.06\pm0.14$ & 11.4  & $-$    \\
UDFz-40566437  &  03:32:40.56  &  -27:46:43.7  &  $27.97\pm0.17$ & $1.32\pm0.40$  &  $-0.07\pm0.15$ &$-0.43\pm0.20$ & 7.7  & $-$    \\
UDFz-41057156  &  03:32:41.05  &  -27:47:15.6  &  $28.08\pm0.16$ & $1.46\pm0.51$  &  $0.18\pm0.16$ &$-0.16\pm0.16$ & 10.7  & $-$    \\
UDFz-36387163  &  03:32:36.38  &  -27:47:16.3  &  $28.27\pm0.18$ & $0.98\pm0.29$  &  $-0.07\pm0.15$ &$-0.12\pm0.17$ & 9.8  & $-$    \\
UDFz-37807405  &  03:32:37.80  &  -27:47:40.5  &  $28.36\pm0.16$ & $>1.97$  &  $0.06\pm0.21$ &$-0.27\pm0.24$ & 6.8  & $-$    \\
UDFz-39736214  &  03:32:39.73  &  -27:46:21.4  &  $28.56\pm0.20$ & $>1.95$  &  $0.04\pm0.23$ &$-0.41\pm0.28$ & 5.9  & $-$    \\
UDFz-38537519  &  03:32:38.53  &  -27:47:51.9  &  $28.96\pm0.24$ & $>1.81$  &  $0.08\pm0.26$ &$-0.62\pm0.36$ & 5.8  & $-$    \\ \hline

UDF-34537360\tablenotemark{a}  &  03:32:34.53  &  -27:47:36.0  &  $25.96\pm0.03$ & $3.35\pm0.57$  &  $0.18\pm0.03$ &$0.21\pm0.03$ & 50.0  &     
\enddata

\tablenotetext{*}{The \jFilter\ photometry was derived from elliptical apertures of 2.5 Kron radii, and has been corrected to total magnitudes by 0.1 mag offsets, while colors were measured in isophotal apertures. Limits correspond to $1\sigma$.}

\tablenotetext{a}{This object is most likely a supernova, as it is very bright and has a stellar profile. It should have been securely detected in the previous NICMOS images, and we therefore exclude it from any further analysis.}

\tablerefs{(1) \citet{bouw06}; (2) \citet{bouw04b}; (3) \citet{yan04}; (4) \citet{coe06}; (5) \citet{labb06}; (6) \citet{bouw08}; (7) \citet{oesch09}; ($\times$) not covered by NICMOS data; ($-$) no previous detection}

\end{deluxetable*}

\section{Observations and source selection}
\label{sec:data}

\subsection{The HUDF09: First Epoch Data}

The HUDF09 program will cover the HUDF \citep{beck06} with a total of 96 orbits of WFC3/IR observations in three filters. These are \hFilter\ (F160W), \jFilter\ (F125W), and \yFilter\ (F105W). The first epoch data available now have been taken between August 26, 2009 and September 6, 2009 and consist of 62 orbits reaching a uniform point-source depth in all filters of $28.6-28.7$ ($5\sigma$ AB, in 0\farcs25 radius apertures).  The dataset covers an area of 4.7 arcmin$^2$ and has a PSF FWHM of $\sim$0\farcs16. It already allows us to generate the deepest NIR images ever seen.

The images have been reduced using standard techniques. A super median image was subtracted from all frames before co-adding them to an image of $0\farcs06$ pixel scale registered to the NICMOS HUDF images \citep[][including the deep UDF05 pointing; see Oesch et al. 2007,2009]{thom05} using our own modified version of \texttt{multidrizzle} \citep{koekemoer02}.
Only ground-based calibration data for WFC3 was available at the time of this analysis. Therefore, image distortion maps have been derived by us based on relatively bright sources in the ACS HUDF. 

We have found that persistence in severely saturated pixels from prior observations can have a noticeable effect on subsequent observations, even hours after the saturation has taken place. The affected pixels have been masked out before further processing. However, two orbits of \yFilter\ data were completely excluded.

We have checked the photometric zeropoints based on SED fitting to the well-calibrated and PSF-matched ACS and NICMOS data, which resulted in values very similar ($<0.05 $ mag) to the space based zeropoints derived by STScI. We therefore opted to use the official values in our subsequent analysis.

\begin{figure*}[tbp]
	\centering
	
		\includegraphics[scale=0.45, angle=-90]{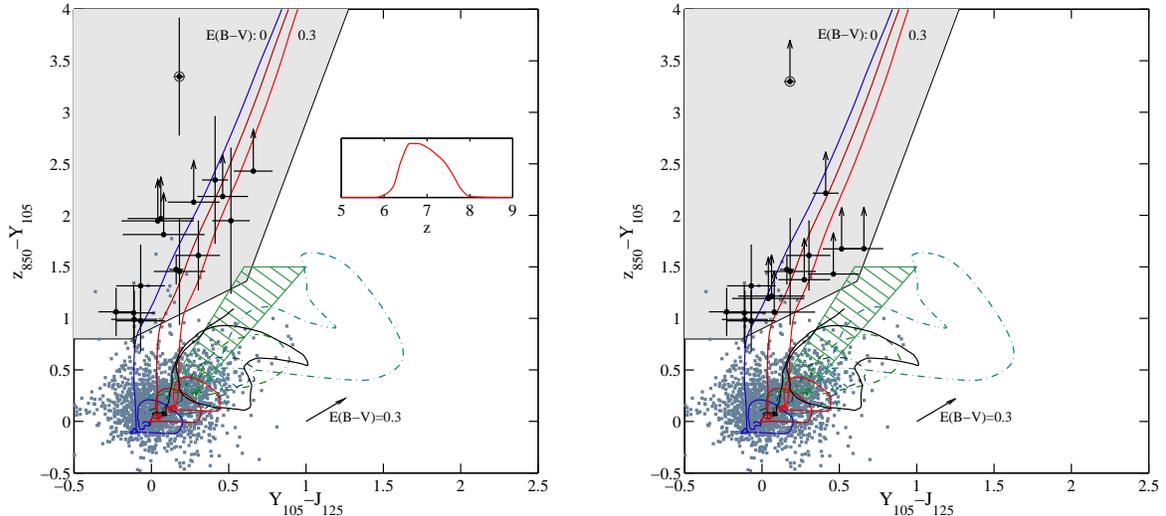}
	\caption{Color-color diagram used for the selection of $z\sim7$ $z_{850}$ dropout galaxies. \textit{(Left)} $1\sigma$ limits are used when measuring colors, \textit{(right)} the same with $2\sigma$ limits. The gray points indicate all galaxies in the catalog and black dots are 17 candidates with no detections in the optical. As mentioned in the text, the reddest of these is very likely to be a SN (marked with a dark gray circle) and is excluded in our further analysis. Tracks of star-forming galaxies, obtained with \citet{bruz03} models, are shown with solid lines in blue-to-red colors corresponding to dust obscurations of $E(B-V)=0,\ 0.15,\ 0.3$. Low redshift galaxy tracks up to redshift $z=5$ are derived from the galaxy templates of \citet{cww80} and are shown as dash-dotted lines. Additionally, we plot the track of a simple stellar population with an age of 500 Myr as a solid black line. The green hatched region covers the locus of M, L, and T dwarf stars derived by convolving observed spectra from \citet{burg04} as well as theoretical template SEDs of \citet{burr06} with the WFC3 filter curves. The black arrow indicates the reddening vector with $E(B-V)=0.3$ at redshift 1.5 and with a \citet{calz00} reddening curve. Low redshift galaxies in our selection window are typically described by very young SEDs with strong nebular line and continuum emission. These galaxies are, however, well detected in the optical data and they all lie within $1\sigma$ of our selection window.
	The inset in the left panel shows the estimated redshift distribution expected for our color selection.}
	\label{fig:colcol}
\end{figure*}

\subsection{\zdrop\ Candidate Selection}
The summed \jFilter+\yFilter\ image is used to detect galaxies with the SExtractor program \citep{bert96} and to measure their colors in isophotal apertures after the optical images have been PSF-matched to the WFC3/IR exposures. Total magnitudes are measured in $2.5$ Kron (AUTO) apertures and small corrections of 0.1 mag have been applied to account for flux loss in the wings of the PSF.

Only sources with signal-to-noise ratio (S/N) larger than 5 within 0\farcs25 radius apertures in both \jFilter- and \yFilter-images are considered.  
We have ensured that our weight maps correctly reproduce the appropriate noise properties in these apertures and measured fluxes $<1\sigma$ were replaced by a 1$-\sigma$ upper limit\footnote{We also checked that the use of 2$-\sigma$ limits does not change our results significantly.  In Fig. \ref{fig:colcol} we show the photometry using both limits.}.

Galaxies are selected from the SExtractor catalog based on the Lyman Break technique \citep[e.g.][]{steidel96,giav04,bouw07} requiring:
\begin{eqnarray*}
	(z_{850}-Y_{105})&>&0.8 \\ 
	(z_{850}-Y_{105})&>&0.9 + 0.75 (Y_{105}-J_{125})  \\
	(z_{850}-Y_{105})&>&-1.1 + 4(Y_{105} - J_{125})  \\
	S/N(J_{125}) > 5 \quad & \wedge & \quad S/N(Y_{105})>5 \\
	S/N(V_{606}) <2 \quad & \wedge & \quad S/N(i_{775}) <2	
\end{eqnarray*}

These criteria select galaxies from $z\sim6.4-7.3$ with a median redshift of $\langle z \rangle = 6.8$, see inset in Fig. \ref{fig:colcol}.

\begin{figure}[tbp]
	\centering
		\includegraphics[width=\linewidth]{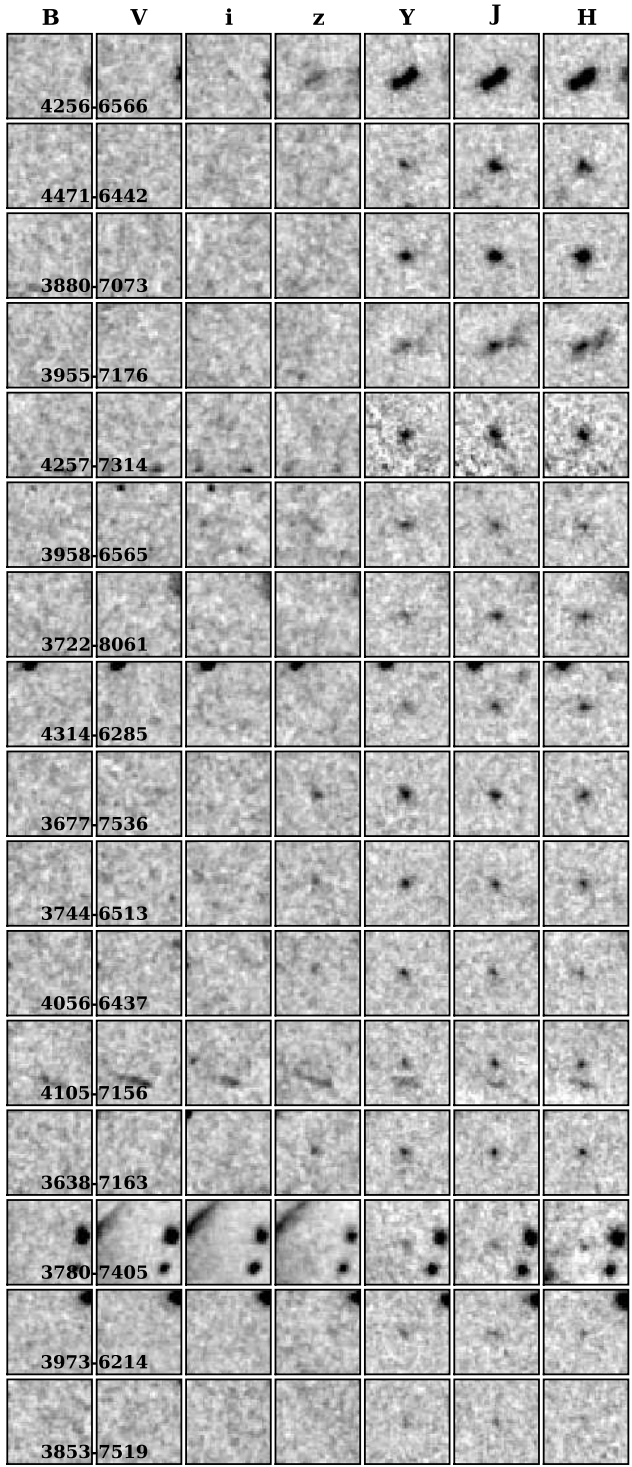}
	\caption{Postage stamps of all $z\sim7$ galaxy candidates in \bFilter, \vFilter, \iFilter, \zFilter, \yFilter, \jFilter, and \hFilter. The sizes of the images are 2\farcs2$\times$ 2\farcs2.
	}
	\label{fig:stamps}
\end{figure}

After rejecting spurious sources such as diffraction spikes of stars and one probable supernova, we find 16 $z\sim7$ candidates between \jFilter$=26.0-29.0$ mag.  Their properties are listed in Table \ref{tab:phot} and an image of all candidates is shown in Fig. \ref{fig:stamps}. 
As can be seen from Table \ref{tab:phot}, the \jh\ color of these sources are extremely blue, in particular towards faint \jFilter\ magnitudes, which results in UV-continuum slopes as steep as $\beta\lesssim-2.5$. This could imply that these sources are dominated by very metal poor stars with little to no dust extinction, or alternatively exhibit initial-mass functions skewed towards massive stars. All these possibilites would have strong implications for the ionizing flux produced by these galaxies, which may be larger than what is assumed in current calculations of reionization \citep[see][]{bouw09d}.

\begin{figure}[tbp]
	\centering
		\includegraphics[scale=0.55]{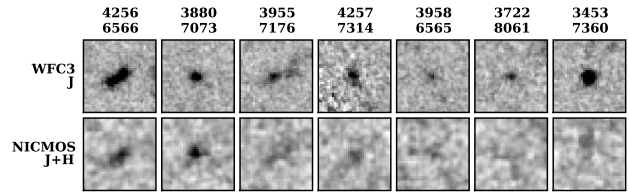}
	\caption{Comparison of WFC3 \jFilter\ (top) and NICMOS F110W+F160W \citep[bottom;][]{thom05} images for a the brightest \zdrop\ galaxies, showing the improvement in data quality. Only galaxies covered with both instruments are shown and only the first four of them have been previously identified as reliable sources. The next two were too faint for detection with NICMOS. The last source is the supernova candidate, which has no significant counterpart in the NICMOS image at its brightness.
	}
	\label{fig:stamps2}
\end{figure}

It is reassuring that all $z\sim7$ galaxy candidates, which have been identified in previous work are confirmed to be secure high redshift candidates (see Table \ref{tab:phot}). 
We show in Fig. \ref{fig:stamps2} a comparison of the NICMOS observations with the WFC3 data for the brightest candidates which were covered by both instruments in order to visualize the enormous improvement in data quality provided by WFC3/IR.
The new data allow us to probe to much fainter limits. While in previous studies only two galaxies were identified beyond a magnitude of 27.5, of which one is only marginally detected \citep{bouw08,oesch09}, the current WFC3 sample includes 11 such faint objects, resulting in much better constraints on the luminosity function at $z\sim7$.

\subsection{Sources of Sample Contamination}

Previous $z\sim7$ selections have suffered from several possible sources of contamination, such as (1) spurious detections, (2) cool dwarf stars, (3) intermediate redshift galaxies with red optical-NIR colors, (4) lower redshift sources which scatter into the selection due to photometric errors, and  (5) high redshift supernovae. Our HUDF09 WFC3/IR observations are much less affected by these problems as we briefly discuss below.

(1) The sources presented in this paper are virtually all $>5\sigma$ detections in three bands, which have been obtained with different dither positions, and the noise properties of WFC3/IR are much better behaved than in NICMOS data. Thus we rule out that any of our source is a spurious detection or is caused by an image artefact\footnote{It is furthermore reassuring that after submission of this manuscript both \citet{mclure09b} and \citet{bunker09} have independently identified a very similar set of $z\sim7$ galaxy candidates from the same dataset used in this work.}.

(2) As can be seen in Fig. \ref{fig:colcol}, dwarf stars occupy a different locus in the \zy\ vs. \yj\ diagram than high redshift galaxies. 
The \jFilter\ band probes short enough wavelengths that it is not dominated by the strong absorption bands of dwarf star SEDs. Therefore, it is very unlikely that any such source contaminates our sample \footnote{Note that the HST \zy\ colors are bluer than $(z-Y)_{AB}$ colors reported for T-dwarfs from ground-based measurements \citep[e.g.][]{pinfield08} due to the wider \yFilter.}.

(3)+(4) The $z\sim7$ galaxy candidates are covered with three bands, all showing colors bluer than expected for possible low redshift contaminants. 
Based on simulations in which we add photometric errors to galaxies in the magnitude range \jFilter$=24-25.9$, simulating the photometry of galaxies down to \jFilter$=29$, we find a negligible contamination fraction by lower redshift interlopers ($<0.3\%$). These simulations are based on the assumption that bright galaxies follow the same SED distribution as at fainter magnitudes, which is not likely to be true. In fact the fainter population shows slightly bluer z-Y colors, which would result in an even smaller contamination rate in the observations than in our simulations. However, the only galaxies with intrinsic SEDs of sufficiently blue NIR colors and which may remain undetected in the optical are very young ($<10^7$ yr) $z\sim1.5$ galaxies with strong emission lines. Given their peculiar SEDs and their extreme faintness being at $z\sim1.5$, such galaxies are not expected to be very common, even though their exact number density is unknown. We are thus confident that our selection is essentially free from low redshift contaminants.

(5) Since our WFC3/IR observations are taken much later than the already existing optical data, supernovae are a potential source of contamination of our sample. Following the calculation in \citet{bouw08}, however, only $0.012$ sources are expected to be found per arcmin$^2$, which results in $\sim0.06$ expected supernovae. Fortunately, at the bright end, such sources can be eliminated by comparison to the existing NICMOS images. Indeed, we find one such source, which shows a stellar profile and, with \jFilter=26.0 mag, should have been securely detected in the previous NICMOS images of the HUDF (see Table \ref{tab:phot}). We exclude this source from our subsequent analysis, but list it here because of its potential interest.

\section{The $z\sim7$ LBG Luminosity Function}
\label{sec:LF}
As in \citet{oesch07,oesch09} completeness, $C$, and magnitude dependent redshift selection probabilities, $S$, for our sample are derived from simulations in which we add artificial galaxies into the real images and rerun our detection and selection procedure. 
The input SED distribution for these simulations are chosen to be slightly steeper than the measured slopes for $z\sim6$ galaxies \citep[$\beta=-2.2$;][]{stan05,bouw09c} to account for the very blue colors seen in our sample ($\beta\sim-2.5$, see also \citep{bouw09d}), and a log-normal size distribution with a size scaling of $(1+z)^{-1}$ \citep{ferg04,bouw04,oesch09b} is assumed.

The luminosity function (LF) brightens significantly from $z\sim6$ to $z\sim4$ \citep[e.g][]{mclure09a,bouw07,bouw06,oesch07,yoshida06,ouchi04,steidel99}, following $M_*(z)=-21.02+0.36*(z-3.8)$ \citep{bouw08}, and previous studies suggested a similar trend to $z\sim7$ \citep[e.g.][]{bouw08,oesch09,ouchi09}. From our simulations we estimate that $31\pm8$ objects should be detected in our small area survey, assuming no evolution of the LF since $z\sim6$ (errorbars include uncertainties in the LF parameters). The cosmic variance is expected to add $35-40\%$ \citep{tre07}. While the expected number of sources is much larger than the 16 observed, the difference is not yet statistically robust. Clearly, the combination of our survey with wider area data will provide much stronger constraints on the changes from $z\sim6$ to $z\sim7$ in the future.

\begin{figure}[tb]
	\centering
		\includegraphics[width=\linewidth]{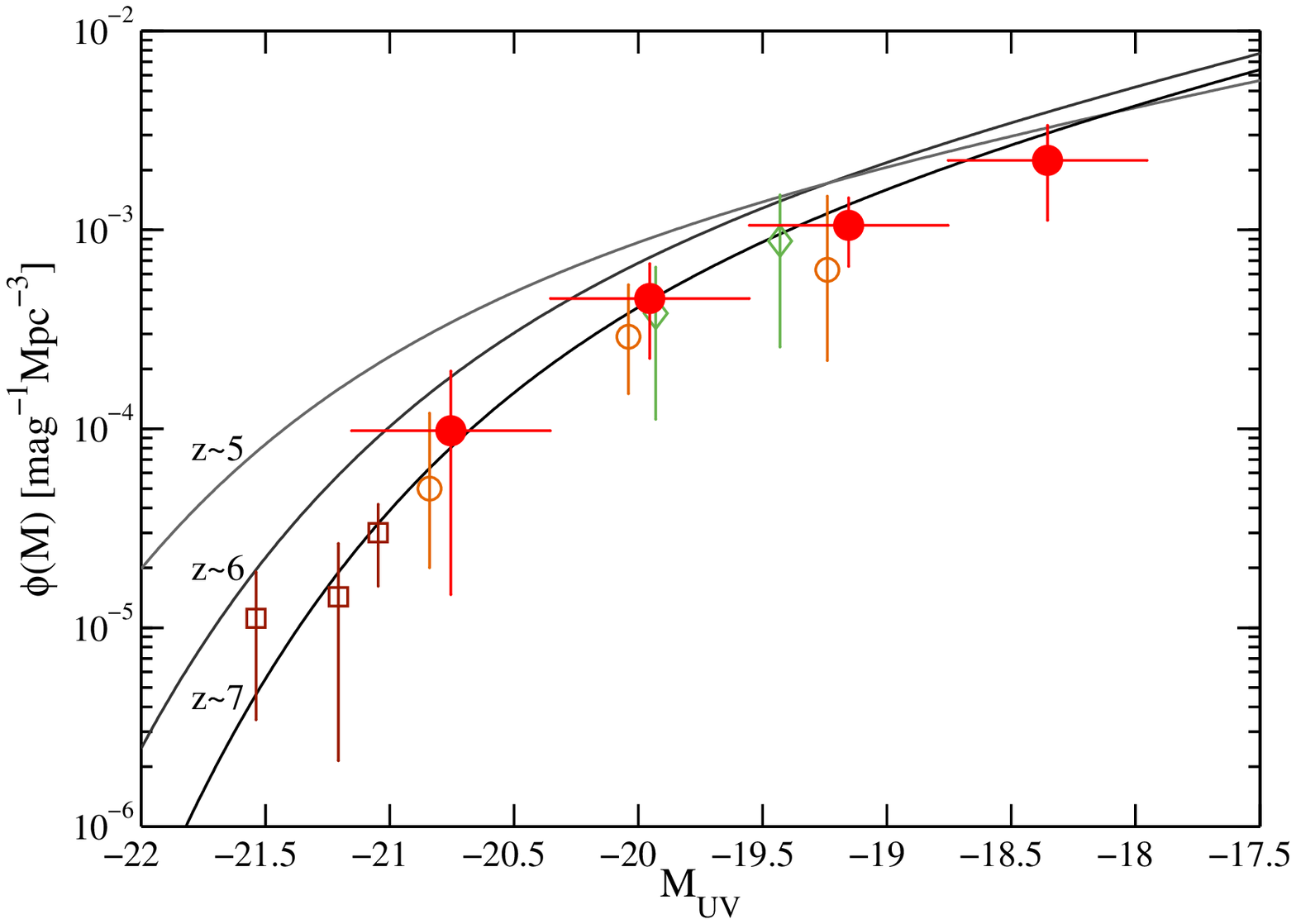}
	\caption{The $z\sim7$ LBG LF derived from the WFC3 candidates (filled red circles). The black solid line corresponds to our best-fit LF with $\alpha=-1.77$, $\phi_*=1.4\times10^{-3}$ Mpc$^{-3}$mag$^{-1}$ and $M_*=-19.91$. 
	%The dashed line is the best-fit for $\phi_*=1.4\times10^{-3}$ Mpc$^{-3}$mag$^{-1}$ and $M_*=-19.8$ resulting in $\alpha=-1.73$. 
	This value of the faint end slope is consistent with no evolution since $z\sim2$ \citep[e.g.][]{bouw07,oesch07,reddy09}. Datapoints from previous estimates are shown as open symbols; orange circles: \citet{bouw08}, green diamonds: \citet{oesch09}, dark red squares: \citet{ouchi09}. The latter ones have been used in our fit to constrain the bright end of the LF. The remaining lines show the LF at $z\sim5$ (light gray solid line; Oesch et al. 2007), and $z\sim6$ (dark gray solid line; Bouwens et al. 2007). Note that the $z\sim7$ points of Bouwens et al. (2008) and Oesch et al. (2009) are partially based on NICMOS data from the HUDF and are thus not independent from our measurements.}
	\label{fig:LFevol7}
\end{figure}

To better quantify the evolution of the LBG population, we show the stepwise LF based on our data in Fig. \ref{fig:LFevol7}. Absolute magnitudes are computed assuming all candidates to lie at $z\sim6.8$ and the effective volume is estimated as $V_{\rm eff}(m) = \int_0^\infty dz \frac{dV}{dz} S(z,m)C(m)$.

Due to the small area of the present data we are not able to fit all parameters of the Schechter function from our dataset alone. We therefore include the points from wide area Subaru searches for $z\sim7$ galaxies by \citet{ouchi09} in order to constrain the bright end of the LF. The fitting is done as in \citet{oesch09} maximizing the Poissonian likelihood for the observed number of sources in each magnitude bin.

The value of $\phi_*$ of the UV LF has been found to be remarkably constant with redshift \citep[see e.g.][]{bouw07}, and we fix its value to $\phi_*=1.4\times10^{-3}$ Mpc$^{-3}$mag$^{-1}$ as measured at $z\sim6$ \citet{bouw07}. This results in $M_*=-19.91\pm0.09$ and $\alpha=-1.77\pm0.20$, which gives a first constraint on the faint-end slope at $z\sim7$. This Schechter function is shown in Fig. \ref{fig:LFevol7} as a black solid line.

The data is consistent with a faint-end slope that is largely unchanged over the 2.5 Gyr from $z\sim7$ to $z\sim2$ with a value of $\alpha\sim-1.7$ \citep[e.g.][]{reddy09,bouw07,oesch07} and the measured value of $M_*$ is in very good agreement with the extrapolation of \citet{bouw08}. Note also that, within the errorbars, the LF function parameters are consistent with the ones derived by \citet{ouchi09}.

\section{Summary and Conclusions}
\label{sec:discussion2}
The first observations with the WFC3/IR camera have demonstrated the amazing improvement over previous NIR instruments. Nevertheless, they have confirmed our prior NICMOS detections at $z\sim7$. In the first epoch data of the HUDF09, which comprises only about a third of the final data set, we have identified a robust sample of 16 $z\sim7$ galaxy candidates down to flux limits as low as $29$ mag. With NICMOS about 100 orbits were needed per $z\sim7$ galaxy candidate found \citep{bouw09a}; WFC3 is $\sim50$ times more efficient and requires only 2.4 orbits per candidate (for fields with existing deep optical data). This remarkable ability to detect high redshift galaxies extends to $z\gtrsim8$ \citep[see][]{bouw09b}.

Most of the $z\sim7$ candidates appear very compact with an average observed half-light radius of $\sim$0\farcs16 (from SExtractor apertures) indicating that the typical size of a starburst galaxy at $z\sim7$ is of order of $\lesssim1$ kpc \citep[for a more extensive analysis of galaxy sizes see][]{oesch09b}.

Due to the depth of our observations we can put a first constraint on the faint end of the LF at $z\sim7$ of $\alpha=-1.77\pm0.20$. This is consistent with no evolution over the whole time span from $z\sim7$ until $z\sim2$ and such a steep faint-end slope has strong implications for reionization, providing support to a scenario in which low luminosity galaxies provide the bulk of the flux for ionizing the universe \citep[see also][]{oesch09,kistler09}.
The combined surface brightness of our sources is $\mu_J \simeq 26.1$ mag arcmin$^{-2}$ which is brighter than the minimum required for reionization at $\sim 7$ for Population II stars $\mu_\mathrm{min} \simeq 27.2$ \citep{stia04,stia04b}. The factor $\sim3$ difference  suggests that even considering an escape fraction $\ll1$, these galaxies still contribute significantly to reionization.

The full HUDF09 data set, complemented with planned wide area surveys, will allow another leap forward in the measurement of the $z\sim7$ LF in the near future. HST is again enabling a revolutionary advance in our understanding of the build-up of galaxies in the universe.

\acknowledgments{We especially thank all those at NASA, STScI and throughout the community
who have worked so diligently to make Hubble the remarkable observatory
that it is today. The servicing missions, like the recent SM4, have
rejuvenated HST and made it an extraordinarily productive scientiﬁc
facility time and time again, and we greatly appreciate the support of
policymakers, and all those in the flight and servicing programs who
contributed to the repeated successes of the HST servicing missions.
PO acknowledges support from the Swiss National Foundation (SNF). 
This work has been supported by NASA grant NAG5-7697 and NASA grant
HST-GO-11563.01.
This research has benefitted from the SpeX Prism Spectral Libraries, maintained by Adam Burgasser at http://www.browndwarfs.org/spexprism.
}

Facilities: \facility{HST(ACS/NICMOS/WFC3)}.

\end{document}